# ULTIMATE FATE OF OUR UNIVERSE FROM QUANTUM MECHANICS

Antonio Alfonso-Faus
E.U.I.T. Aeronáutica, Plaza Cardenal Cisneros s/n
28040 Madrid, SPAIN

e-mail: antonio.alfonso@upm.es

Abstract

It is conjectured that time intervals of any kind are proportional to the age of the Universe taken at the time we are considering the interval. If this is the case then the speed of light, in fact any speed, must decrease inversely proportional to this age. The immediate consequence is that energy is not conserved: the hypothesis that time is a homogeneous property implies conservation of energy (the theorem of Noether). Non-conservation of energy follows from the condition that any time interval is proportional to the cosmological time, and therefore time can not be homogeneous. From the uncertainty principle, taking the constant of Planck as a real constant, time independent, it follows that any energy in the Universe decreases linearly with time. We then prove that Schroedinger equation does not change, except for the potential energy term. The future of the Universe gives for the wave functions a long sinusoidal spatial solution, so that everything becomes unlocalized. The relativistic absolute interval remains the same, even with a changing speed of light, and the Universe turns out to be non-expanding. A Mass-Boom effect is confirmed.

The only way we have to measure the physical property we call time is by counting oscillations. A clock, the device to measure time, is an oscillating system and the numerical sequence of these oscillations gives us a reference number to fix an event at a certain "time". Now, do clocks have a homogeneous tic? The time intervals represented by the individual tics may be all equal making us to think that time is homogeneous, and consequently that energy is a conserved quantity. But given a clock, how do we know that it is ticking homogeneously? No way out. We can only compare clocks, and we never know if there is anyone ticking in a homogeneous way. Consider a fundamental particle, say a proton, of mass m. Its Compton size is $\hbar/mc$, where $\hbar$ is Planck's constant and c the speed of light. Taking Planck's constant as a real time independent constant and the product mc as representative of momentum, a conserved quantity if space is homogeneous, and then the size of the particle is a constant. The time light takes to travel this size is then $\hbar/mc^2 = \tau$. This basic time interval is inversely proportional to the speed of light. Now, there is good evidence [1] for the speed of light to be time varying, inversely proportional to the cosmological time t. Then the time interval given by one tic of an atomic clock, which has tics proportional to $\tau$, is also proportional to the age of the Universe t. Gravitational clocks can be shown to have the same ticking property. Then it is a plausible conjecture that all clocks, good clocks, are slowing down because their tics are increasing in length proportionally to t.

We have now a very interesting property of time measuring. If our conjecture is true then there is a constant number of tics at any age of the Universe t. if we divide the age t by the time interval between tics, which are proportional to t, then we get the constant number $\cong 10^{41}$. This is in contrast with Dirac's [2] approach that takes the tics as homogeneous, constant, and interprets the number $10^{41}$ as representative of the age of the Universe, the cosmological time t. Our conjecture makes this number constant.

On the other hand we can count the number of fundamental clocks of this kind, in the visible Universe, as $10^{80}$, which is also a constant. Multiplying both numbers we get the very well known number in cosmology $10^{121}$. Here we interpret this constant number as the total number of tics in the Universe at any age. The time interval $t/10^{121}$ is the basic tic of the visible Universe at age t, which is of course proportional to t, in accordance with our conjecture. This basic tic is also equal to $\hbar/Mc^2$, where M is the mass of the visible Universe $M = 10^{80}$ m.

Now, can we say that the age of the Universe, or the universal time t, is a continuous property (general relativity) or is it a discrete property (quantum mechanics)? We believe that the answer is both at the same time, like any other physical property in nature, as it corresponds to the dual nature (particle and wave) of everything. The time interval $t/10^{121}$ corresponds to a frequency $\upsilon = 2\pi 10^{121}/t$, and therefore the total energy in the Universe is

$$E = \hbar \upsilon = 2\pi 10^{121} \frac{\hbar}{t} \tag{1}$$

We know that the mass of the quantum of gravity $m_g$ is [3]

$$m_g = \frac{\hbar}{c^2 t} = 2.10^{-66} \, grams \tag{2}$$

therefore the energy of the quantum of gravity is

$$m_g c^2 = \frac{\hbar}{t} = 2.10^{-45} \, ergs \tag{3}$$

and from (1) we see that in the Universe there are $2\pi 10^{121}$ gravity quanta, a fixed constant number. From (3) we also recognize that the Compton wavelength of the quantum of gravity $\hbar/m_g c$ is ct, i.e., the size of the visible Universe. This means that the uncertainty in the position of the quantum of gravity is just the size of the Universe, and therefore is unlocalized as it corresponds to a gravitational energy, the energy of the gravitational field [4].

The immediate consequence of the outmost importance is that energy decreases with cosmological time as 1/t. Since any energy is the product of momentum times the speed of light, or any speed v proportional to c (to preserve relativity v/c must be constant), then from the constancy of momentum (the most conserved property in nature) we get that the speed of light, in fact any speed, must decrease as 1/t. Then the product ct is a constant and the visible Universe DOES NOT EXPAND [5]. The observed Hubble red shift must be explained by a cause different from expansion. Then the wavelength of the quantum of gravity is also a constant. It also propagates at the speed of light, as any physical interaction.

Separating the wave function in a product of two factors, the spatial part and the temporal one, one gets the time independent Schrödinger equation [7]. To be a real time independent wave equation it must be multiplied by the age of the Universe t. This is because the Schrödinger equation has its terms with dimension of energy, and therefore both members of the equation decrease with time. To convert it in a real time independent equation a factor of t (in fact $t/t_0$, dimensionless with $t_0$ a reference age like today's age) must be multiplying both sides. The final Schrödinger equation reads then as follows:

$$\frac{1}{\psi(x)}\left(-\frac{\hbar^2}{2}\frac{1}{m}\frac{t}{t_0}\frac{d^2\psi(x)}{dx^2}+\frac{t}{t_0}V(x)\psi(x)\right)=i\hbar\frac{1}{\varphi(t)}\frac{t}{t_0}\frac{d\varphi(t)}{dt}=const.=\frac{Et}{t_0} \quad (4)$$

The effect of this factor of t is enormous. First, the constancy of momentum together with the decrease of the speed of light implies a linear increase of mass with time t (the Mass-Boom effect presented elsewhere [6]). Hence the ratio t/m in (4) is a constant. The conclusion is that the time independent Schrödinger equation, the spatial part in x, the first term in mass does not change at all and remains as usual. The only thing is that the mass m has to be referred to the rest of the mass of the Universe, m/t being this constant proportion. And this is a Machean result introduced into the Schrödinger equation, a cosmological parameter introduced into the quantum mechanical approach. The second term in the spatial part in x, the potential energy V(x) has to be substituted by $V(x).t/t_0$. This is a very important change. Since we have the equality $V(x).t = V_0(x).t_0$ this means that the potential energy in the usual spatial part of Schrödinger equation has to be $V_0(x)$, i.e. time independent. It is the value at age $t_0$ and has to be kept constant. In other words, the time variation of the potential does not have to appear explicitly in the equation, and the value of the potential is the one at age $t_0$, no more (before) no less (afterwards). Hence, the application of Schrödinger equation to any problem in quantum mechanics is valid only for short time intervals around the time we are considering the "present", the age at which we are applying the equation. The equation changes for different times in a cosmological sense. And therefore the spatial part of the equation is in fact time dependent in a cosmological sense.

Given that the energy decreases with time, the potential energy in the far future tends to zero. The Machean approach that ascribes the relativistic energy of any particle $mc^2$ to the gravitational potential energy of m with respect to the rest of the Universe implies that the future of any energy is zero. And no potential energy means no confinement also, and therefore that the wave function spreads and everything becomes "unlocalized". This is the ultimate fate of the Universe.

The integration of the second member of (4) is straightforward from the transformed:

$$\frac{d \ln \varphi(t)}{d \ln t} = const. = -i\frac{Et}{\hbar} \qquad (5)$$

i.e.

$$\ln \varphi(t) = -i\frac{Et}{\hbar} \ln At \qquad (6)$$

where A is a constant. Therefore we get for the time part of the wave function

$$\varphi(t) = const. \left(\frac{t}{t_0}\right)^{-i\frac{Et}{\hbar}} = const. \, e^{-i\frac{Et}{\hbar} \ln \frac{t}{t_0}} \qquad (7)$$

Now, if t is close to $t_0$ then ln (t/$t_0$) is close to (t – $t_0$)/$t_0$ which is equal to δt/$t_0$ ( δt being a small time interval after the age $t_0$. The time part of the wave function is finally (with Et = $E_0 t_0$)

$$\varphi(t) = const. \, e^{-i\frac{E_0 \partial t}{\hbar}} \qquad (8)$$

which is the usual one. No change. The conclusions are then as follows:

1) The conjecture that any time interval, any "tic" from any clock, is proportional to the age of the Universe implies that the speed of light, any speed, must decrease inversely proportional to cosmological time.

2) There is a Mass-Boom effect following conservation of momentum. All gravitational masses increase linearly with cosmological time. They do so creating in this way the gravitational quanta, the gravitational field.

3) The Universe is not expanding. There are many alternatives to explain the Hubble red shift.

4) The ultimate fate for anything in the Universe, the future, is a long sinusoidal wave function so that everything becomes unlocalized, fading away.

5) Finally, the absolute General Relativity interval ds has a first term $(cdt)^2$ that in our case, given our conjecture is c inversely proportional to t, and dt being proportional to age t, dt = tdt', results in the expression $(cdt)^2$ = constant . $(dt')^2$. Also a non-expanding Universe means a cosmological scale factor R = constant. Then the local absolute interval is just the Minkowski interval with c = 1 in a certain system of units, and the General Relativity ds has also constant coefficients in front of the coordinate differentials. To all intents and purposes our treatment leaves the intervals as if c=1 and the Universe were not expanding, R = 1.

REFERENCES

[1] Yves-Henri Sanejouand, arXiv:astro-ph/0509562 v1 20 Sep 2005

[2] Dirac, P.A.M., Nature, **139**, 323 (1937).

[3] Alfonso-Faus, A., Physics Essays, Vol. 12, N4, December 2000.

[4] Misner, C.W., Thorne, K.S. and Wheeler, J.A. in Gravitation,1973, Freeman & Co.

[5] Alfonso-Faus, A., arXiv:physics/0701104 and AIP Book of Proceedings, 8[th] International Symposium "Frontiers of Fundamental Physics", Madrid 2006.

[6] Eisberg, R.M., Fundamentals of Modern Physics, 1961, John Wiley & Sons.

[7] Alfonso-Faus, A., AIP Book of Proceedings, 1[st] Crisis in Cosmology Conference, Monçāo, Portugal 2005.